\documentclass[prb,twocolumn,superscriptaddress,showpacs,amssymb]{revtex4-1}

\usepackage{standalone}
\usepackage{soul}
\usepackage{graphicx}
\usepackage{color}

\begin{document}
\title{Interaction distance  in the extended XXZ model}

\author{Kristian Patrick}
\affiliation{School of Physics and Astronomy, University of Leeds, Leeds, LS2 9JT, United Kingdom}
\author{Vincent Caudrelier}
\affiliation{School of Mathematics, University of Leeds, Leeds, LS2 9JT, United Kingdom}
\author{Zlatko Papi\'c}
\affiliation{School of Physics and Astronomy, University of Leeds, Leeds, LS2 9JT, United Kingdom}
\author{Jiannis K. Pachos}
\affiliation{School of Physics and Astronomy, University of Leeds, Leeds, LS2 9JT, United Kingdom}

\date{\today}

\begin{abstract}
We employ  the interaction distance to characterise the physics of a one-dimensional extended XXZ spin model, whose phase diagram consists of both integrable and non-integrable regimes, with various types of ordering, e.g., a gapless Luttinger liquid and gapped crystalline phases. We numerically demonstrate that the interaction distance successfully reveals the known behaviour of the model in its integrable regime. As an additional diagnostic tool, we introduce  the notion of ``integrability distance"  and particularise it to the XXZ model in order to quantity how far the ground state of the extended XXZ model is from being integrable. This distance provides insight into the properties of the gapless Luttinger liquid phase in the presence of next-nearest neighbour spin interactions which break integrability.

\end{abstract}

\maketitle

\section{Introduction}

An efficient way for describing a many-body quantum system is by identifying the effective degrees of freedom that encapsulate its dominant  low energy properties~\cite{LandauLifshitz}.
In a similar vein to Fermi liquid theory, which applies to weakly correlated systems where the effective degrees of freedom are ``dressed" versions of the original degrees of freedom, it would be desirable to have general techniques to characterise the effect of interactions in general (possibly strongly-correlated models), without relying on the specific physics or \emph{exact} mathematical structure (e.g., integrability~\cite{Sutherland}) of the model. 
The \emph{interaction distance}~\cite{Turner} provides a systematic measure of the effect interactions can have on a given quantum state of a generic many-body system. 
In the case of a reduced density matrix (see section \ref{sec:df}), the interaction distance is determined solely from the entanglement spectrum~\cite{Haldane08} of the given quantum state; intuitively, it captures the long distance behaviour of a system by identifying the quantum correlations between the emerging degrees of freedom. At the same time, it includes information about the structure of the degrees of freedom that are dressed by the interactions, thus revealing the short distance behaviour of the model. 

The interaction distance compares the correlations of a system to those of chosen free particles, which we assume in this work to be fermions. Moreover, it identifies the optimal free model closest to the interacting one, thus offering a qualitative and quantitative analysis of the interacting system. Free fermions are a restricted class of models that are analytically tractable; to accommodate a more general class, below we introduce the notion of ``integrability distance'' -- a distance that measures how far the correlations of the given state of a generic interacting system are from the closest possible {\em integrable} model. This measure allows to identify  in principle if a certain physical model is ``almost" integrable, thus potentially rendering it amenable to some of the analytical tools of integrability. 

We exemplify our approach using the extended XXZ model -- a non-integrable model hosting a Luttinger liquid phase~\cite{Tomonaga, Luttinger} -- that has been under intense investigation in recent years. In the integrable limit, properties of the model have been studied extensively via Bethe ansatz~\cite{Gaudin},  leading to exact results for correlation functions both analytically~\cite{korepin_bogoliubov_izergin_1993} and numerically\cite{Caux}, and density matrix renormalisation group (DMRG)~\cite{White}, either in finite~\cite{Schollwock} or infinite systems~\cite{VidaliDMRG,McCulloch}. In particular, properties of the gapless Luttinger liquid phase have been tested via explicit calculations of equal-time density response functions~\cite{Schmitteckert,Ejima}, the density of states at zero temperature~\cite{Pereira,Pereira2,Jeckelmann}, and power-law decay of correlation functions~\cite{Karrasch}. Furthermore, it was shown that the Luttinger liquid phase remains stable to a small amount of integrability-breaking next-nearest neighbour interactions~\cite{Zhuravlev,SchmitteckertExt,Duan}. More recently, extensive DMRG studies have mapped out the phase diagram of the extended XXZ model~\cite{Mishra}, and Luttinger liquid physics has been probed via quantum quenches~\cite{Dora,KarraschQuench}.

Here, we demonstrate that a scaling analysis of the interaction distance successfully reproduces the asymptotic free-fermion behaviour of the extended XXZ model when restricted to its gapped integrable regime. Moreover, we employ the integrability distance in order to investigate the non-integrable version of the model obtained by including longer-range interactions between spins. As the integrability distance is too complex to determine in its full generality, we present a physically motivated simplified procedure that is suitable for describing the  extended XXZ model. This distance measures how faithfully the entanglement properties of the ground state of the (non-integrable) extended XXZ model can be represented by the Luttinger liquid, thus allowing for a quantitative understanding of the extended model and potentially tractable analytic treatment. 

The paper is organised as follows. In Section II we provide a brief overview of interaction distance and discuss the physical meaning of this quantity, in particular how it can probe both short and long distance behaviour of the model. In Section III we introduce the extended XXZ model that we employ to demonstrate the diagnostic ability of the interaction distance. Section IV analyses the integrable XXZ model in terms of the interaction distance. In this section we identify the asymptotically free behaviour of the model in its gapped region and perform a non-perturbative calculation of the interaction distance for the gapless regime.  Section V presents the analysis of the extended XXZ model from the perspective of  the ``integrability distance", that is introduced in order to measure the closeness of the correlations in the ground state of the extended model to the correlations in the integrable regime. Our conclusions and outlook are presented in Section VI.

\section{Quantifying the effect of interactions}\label{sec:df}

This section provides a self-contained overview of  the interaction distance, $D_{\cal F}$, that was originally introduced in Ref.~\onlinecite{Turner} (see also Ref.~\onlinecite{PachosScipost}).  The interaction distance is the tool we use in this paper in order to quantify the effect of interactions on a quantum system. Intuitively, we expect a quantum system to be ``non-interacting" if we are able to express its Hamiltonian in a quadratic form, in terms of some suitably defined creation and annihilation operators. However, this requirement may be too stringent for many purposes where the main focus is only on the ground state of the system and a few low-lying excited states. In such cases, we are motivated to redefine ``freedom" with respect to the given quantum state, or more precisely its reduced density matrix being approximately expressible in a quadratic form.  

\subsection{Interaction distance}

We focus on lattice models of interacting fermions, described by creation and annihilation operators, $c_j^\dagger$, $c_j$. The interaction distance  $D_{\cal F}$ is defined as the trace distance between the density matrix $\rho$ of an arbitrary quantum system and the closest density matrix corresponding to some free system $\sigma$, given by~\cite{Turner}
\begin{eqnarray}
D_{\cal F}(\rho) = \min_{\sigma \in {\cal F}} {1 \over 2}\mathrm{tr}\left( \sqrt{(\rho-\sigma)^2}\right).
\label{eqn:intdist}
\end{eqnarray}
The minimisation is performed over all free density matrices $\sigma$, which belong to the manifold of Gaussian (free) fermion states $\mathcal{F}$. Specifically, we can write
\begin{eqnarray}\label{eq:sigma}
\sigma = \frac{1}{Z_\sigma} \exp\left(-\beta \sum_j \epsilon_j f_j^\dagger f_j^{} \right),
\end{eqnarray}
where $f_j$ are some fermion operators, $\beta$ denotes inverse temperature, and $Z_\sigma$ is a normalisation constant which ensures $\mathrm{tr}\;\sigma=1$ (we also assume that $\rho$ is normalised in the same manner). Note that $f_j$ are not necessarily the same as the original fermionic operators $c_j$ that appear in the Hamiltonian describing the system. Moreover, we emphasise that the trace distance is merely one convenient choice for the definition of $D_{\cal F}$, and other choices like relative entropy~\cite{NielsenChuang} can equally well be used.

Expressions Eqs.~(\ref{eqn:intdist})-(\ref{eq:sigma}) can be used in formally the same way in two very different physical contexts: $\rho$ can represent the Boltzmann-Gibbs density matrix of the system, or it can be a \emph{reduced} density matrix, which describes a subsystem $A$ for some (real space) partition of the total system in $A$ and its complement $B$. In the latter case, assuming that the entire system is in a pure state $|\psi\rangle$, the reduced density matrix $\rho_A$ is defined as
\begin{eqnarray}
\rho_A = \mathrm{tr}_B |\psi\rangle\langle\psi|,
\end{eqnarray}
where $\mathrm{tr}_B$ denotes the partial trace over the degrees of freedom in $B$. In general, the reduced density matrix $\rho_A$ describes a mixed state, with some effective temperature $\beta=1$. The negative logarithm of the eigenvalues of $\rho_A$, i.e., $-\ln\rho_{k}$, is known as the ``entanglement spectrum"~\cite{Haldane08}. In the case of systems with conformal invariance~\cite{Casini2011,CardyTonni}  or in topological phases of matter~\cite{QiKatsuraLudwig}, the entanglement spectrum inherits some characteristics of the energy spectrum of the full system, e.g., it reveals the energy excitations at the edge of a topologically ordered system~\cite{Haldane08}.  In this paper, we focus on the reduced density matrix case of $D_{\cal F}$ for reasons explained in Section \ref{reasons}.

We note that the definition of $D_{\cal F}$ in Eq.~(\ref{eqn:intdist}) appears to require a difficult minimisation over all $\sigma\in\mathcal{F}$. Nevertheless, it has been shown that the minimum value can be computed simply from the spectra of $\rho$ and $\sigma$~\cite{Markham, Turner}. Taking this into account, 
the interaction distance is equivalently given by the simpler expression
\begin{eqnarray}
D_{\cal F}(\rho) = \min_{\{\epsilon_j\}} {1 \over 2} \sum_k \left| e^{- \beta E_k} -e^{- \beta E_k^\mathrm{f}(\epsilon)} \right|,
\label{eqn:intdis_2}
\end{eqnarray}
where we have introduced the notation $e^{-\beta E_k}$ for the $k$th eigenvalue of $\rho$, and the eigenvalues of $\sigma$ are similarly given in terms of
\begin{eqnarray}
E_k^\mathrm{f}(\epsilon) = E_0 +  \sum_{j}  \epsilon_j n_j^{(k)}.
\label{eqn:freespec}
\end{eqnarray}
For every $k$ in Eq.~(\ref{eqn:freespec}) there is a specific pattern of fermionic populations $n_j$'s that take values 0 or 1, and $E_0$ guarantees the normalisation of $\sigma$. 

The advantage of Eq.~(\ref{eqn:intdis_2}) is that the minimisation is only with respect to the single particle energies $\{ \epsilon_j \}$, whose number typically scales linearly with the total size of the system. This is in contrast to the total number of eigenvalues, whose number is exponential in the size of the system or subsystem, depending on whether $\rho$ is a thermal or reduced density matrix. Thus, $D_{\cal F}$ is a diagnostic tool that can be efficiently computed numerically or analytically for any system whenever its energy or entanglement spectrum $\{E_k\}$ is accessible. 

\subsection{Short and long distance behaviours}\label{reasons}

The interaction distance $D_{\cal F}$ in Eq.~(\ref{eqn:intdist}) expresses the distinguishability~\cite{NielsenChuang} of the two density matrices, $\rho$ and $\sigma$. It has a geometric interpretation as the distance of the density matrix $\rho$ from the manifold ${\cal F}$~\cite{PachosScipost}. Importantly, the optimal free state $\sigma$, i.e., the one with the smallest distance from $\rho$, does \emph{not} need to be expressed in terms of the original degrees of freedom, $c_j$, that define the Hamiltonian. Moreover, in general the optimal free state may not be unique, although in many cases it was indeed found to be~\cite{Turner, Hills}.
When $\rho$ is chosen to be the reduced density matrix, $D_{\cal F}$ measures the distance of the entanglement spectrum, corresponding to the given state $|\psi\rangle$ and the given partition, from the closest possible free-fermion entanglement spectrum, $\{E_k^\mathrm{f}\}$, given by Eq.~(\ref{eqn:freespec}). Loosely speaking, $D_{\cal F}$ measures how much the part $A$ of the system ``interacts" with part $B$~\cite{Genoni,ivan2012,AdessoGaussian,Marian,GaussianRMP,Gertis}.

An important characteristic of $D_{\cal F}$ is that it explicitly depends on the partition between $A$ and $B$ subsystems, which can impact its behaviour. In the majority of physical systems, the low-energy physics can be described in terms of weakly interacting degrees of freedom (DoF), which are expressible in terms of dressed  original degrees of freedom, $U c_j U^\dagger$, where $U$ is some unitary transformation. As $U c_j U^\dagger$ is a canonical transformation, the resulting operators are still fermionic, but with possibly different characteristics. For example, they might have support on a larger region than just one site $j$ depending on the action of $U$ on them. 

\begin{figure}[thb]
\begin{center}
\includegraphics[width=\linewidth]{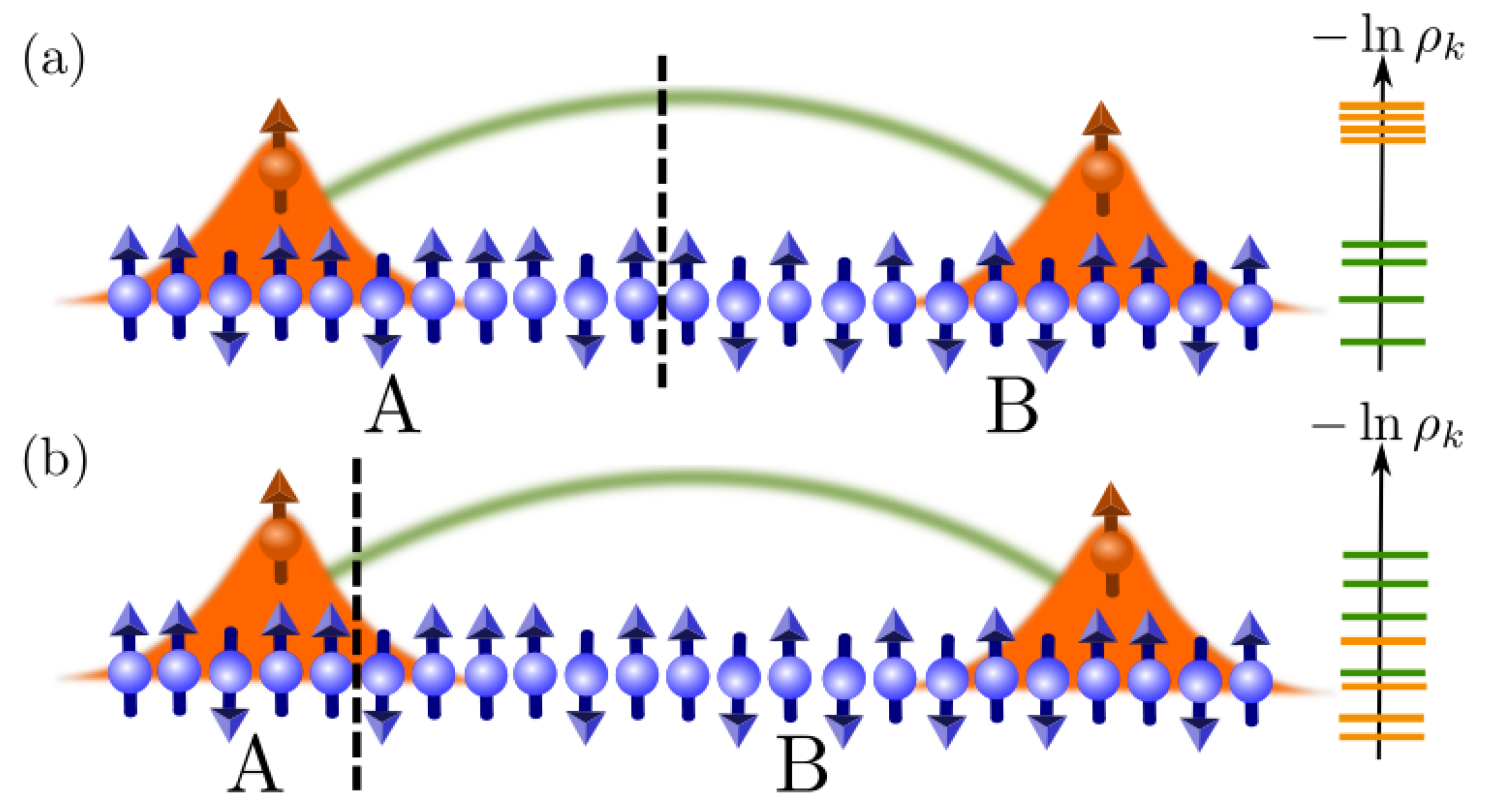}
\end{center}
\caption{(a) The dressed degrees of freedom (orange) that effectively describe the behaviour of the interacting system are much smaller than the length, $L_A$ of the subsystem. The low-lying entanglement spectrum is universal and describes the correlations (green line) between such dressed DoF~\cite{Haldane08} on either side of the partition (black dashed line). The orange part of the entanglement spectrum is associated with the structure of the dressed DoF and it is expected to be separated from the universal green part through an ``entanglement gap". (b) The size of the dressed DoF is comparable to the length of the subsystem, which means that partitioning the system would necessarily ``cut" through a DoF. In this case, the low-lying entanglement spectrum also probes the internal structure of the dressed DoF.
}
\label{fig:qp}
\end{figure}

In cases that are amenable to mean-field theory, $U$ can be decomposed into \emph{linear} transformations of $c_j$ operators. In this case the dressed DoF of the system are the initial fermions, $c_j$, while the linear transformation determines the quantum correlations between them. However, more complicated choices of $U$ are also possible, which cannot be expressed as linear transformations of the original degrees of freedom. Such operators create more complicated types of fermionic DoF that have non-trivial internal structure and may be supported over a larger range of lattice sites, an example is sketched in Fig.~\ref{fig:qp}. In all cases, the correlations between such DoF underpin the low-energy properties of the model. We note that similar ideas have recently been used in the framework of matrix product state methods in order to construct quasiparticle excitations in various 1D models~\cite{Smatrix,Scattering,VanderstraetenThesis}. 

If the size of the subsystem $A$ is much larger than the typical size $\ell$ of the dressed DoF (given by the spatial support of the operator $U$ acting on $c_j$'s) as in Fig.~\ref{fig:qp}(a), then the bottom part of the entanglement spectrum captures the correlations between the dressed DoF. The top part of the entanglement spectrum encodes the structure of the dressed DoF isolated from the bottom part by the existence of an ``entanglement gap"~\cite{Haldane08}. As seen from (\ref{eqn:intdis_2}) the contributions from the top part of the entanglement spectrum to the interaction distance is exponentially suppressed. Hence, in this case $D_{\cal F}$ probes the large distance behaviour, i.e. the correlations between dressed DoF.

If, on the other hand, $\ell$ is comparable with the size $L_A$ of the subsystem $A$ this implies that the entanglement partition will necessarily split the dressed DoF. Hence, the interaction distance will probe the physics associated with their internal structure, i.e. short distance behaviour. 
This is likely to happen at critical points and second-order phase transitions, where the size, $\ell$, may diverge. As a result, the behaviour of $D_{\cal F} $ in the two cases may be different, probing the long distance or the short distance behaviour of the system, depending on its critical behaviour or the size of the partition. 

From the above discussion it becomes apparent that $D_{\cal F}$ quantifies the non-linear effect interactions may have on the DoF of a system. This distinguishes the interaction distance from other diagnostic tools such as two-point correlations, where the linear and non-linear contributions in general both contribute. To illustrate this, note that for the XY spin model (which is a special limit of the extended XXZ model, as discussed below), the spin-spin correlation function is given by~\cite{LSM}
\begin{eqnarray}
\langle S_0^z S_n^z \rangle = -\frac{1}{4}\left( \frac{2}{n\pi}\right)^2,
\end{eqnarray}
which is valid for $n$-odd. At the same time, the XY model can be easily diagonalised by performing the Fourier transformation, which maps it to free fermions in momentum space. Consequently, we obtain $D_\mathcal{F}=0$ for the XY model for any choice of the partition. This example illustrates that $D_\mathcal{F}$ only captures the non-linear part of the correlations between the original degrees of freedom.

\section{The extended XXZ model}\label{sec:ext}

In oder to systematically investigate the effect interactions can have on the low-energy spectrum we employ a specific example. The systems we have in mind are defined on a lattice, e.g., a system of quantum spins with a local Hilbert space and local (nearest neighbour) hopping and interaction terms. 
For concreteness, we focus on the extended XXZ spin-$1/2$ model, described by the Hamiltonian
\begin{eqnarray}\label{eq:xxz}
\nonumber H_{XXZ} &=& J \sum_j \left( S_j^x S_{j+1}^x + S_j^y S_{j+1}^y  \right) + J_{zz} \sum_j S_j^z S_{j+1}^z \\
&+& J_{zz}' \sum_j S_j^z S_{j+2}^z,
\end{eqnarray}
where $S_j^\alpha$ are the standard spin-$1/2$ operators on site $j$, $J$ is the hopping amplitude (we set $J=1$), and we included interactions between nearest-neighbour spins ($J_{zz}$) as well as between next-nearest neighbours ($J_{zz}'$). 

For a one-dimensional system like in Eq.~(\ref{eq:xxz}),  in the case $J_{zz}'=0$, the effect of nearest-neighbour interactions can be rigorously accounted for via integrability techniques~\cite{baxter,Gaudin} (in particular, algebraic/coordinate Bethe ansatz) for arbitrary values of $J_{zz}$. However, integrability is broken as soon as we include interactions between next-nearest neighbour spins ($J_{zz}'$), or by generalising the model to higher dimensions. On the other hand, techniques such as bosonisation~\cite{BosonizationBook} are very versatile at describing a large class of \emph{gapless} systems that behave as Luttinger liquids (LLs). For the model in Eq.~(\ref{eq:xxz}) in the absence of $J_{zz}'$ term, the LL phase occurs for $|J_{zz}|<1$~\cite{giamarchi2004quantum}. Numerical studies using DMRG~\cite{Mishra} have shown that the LL phase survives in a finite range of $J_{zz}'>0$, and is surrounded by two types of charge-density wave phases and a bond-ordered phase. 

We also remind the reader that the one-dimensional model in Eq.~(\ref{eq:xxz})
can be directly recast via Jordan-Wigner transformation as a system of  spinless fermions hopping on a lattice, 
\begin{eqnarray}\label{eq:xxz_2}
\nonumber H_{XXZ} &=& \frac{J}{2} \sum_j \left( c_j^\dagger c_{j+1} + \mathrm{h.c.} \right) \\
\nonumber &+& J_{zz} \sum_j \left(n_j - \frac{1}{2}\right) \left(n_{j+1} - \frac{1}{2}\right) \\
&+& J_{zz}' \sum_j \left(n_j - \frac{1}{2}\right) \left(n_{j+2} - \frac{1}{2}\right),
\end{eqnarray}
with nearest neighbour and next-nearest neighbour density-density interactions ($n_j \equiv c_j^\dagger c_j$). With antiferromagnetic $J_{zz}>0$ (and $J_{zz}'=0$), Eq.~(\ref{eq:xxz_2}) captures the low-energy physics of the 1D Fermi-Hubbard model at large interaction $U$~\cite{Korepin}. Thus, even in simple models like in Eq.~(\ref{eq:xxz}) or Eq.~(\ref{eq:xxz_2}), we see that the effects of interactions can be very complex, and lead to a variety of behaviours (gapped or gapless, integrable or non-integrable, etc.). In the following we employ the interaction distance and the new concept of integrability distance to numerically investigate the low-energy properties of this system. For smaller system sizes, we use periodic boundary conditions and obtain the ground state numerically using exact diagonalisation, resolving the translation symmetry of the system. Alternatively, to access larger system sizes, we assume open boundary conditions and use DMRG method, implemented in ITensor~\cite{ITensor}, to variationally obtain the ground state of the system and its entanglement spectrum. Unless specified otherwise, the entanglement spectrum is obtained by partitioning the system in real space in two subsystems of equal size. As we explained in Sec.~\ref{sec:df}, from the knowledge of the entanglement spectrum, we can efficiently evaluate $D_{\cal F} $.

\section{Quantifying interactions in the integrable XXZ model}

Here we consider the integrable part of $H_{XXZ}$ with $J_{zz}'=0$. Our aim is to establish how well the interaction distance, $D_{\cal F} $, can capture the behaviour of the model known from its analytical treatment. 

\subsection{Gapped antiferromagnetic phase of XXZ model}

First, we turn our attention to the gapped phase of the integrable XXZ model, i.e. with $J_{zz}>1$. In Refs.~\onlinecite{PeschelIntegrable,Alba2017} it was shown that the reduced density matrix of an infinite system, bipartitioned into two semi-infinite lines, can be written as 
\begin{eqnarray}\label{eq:gapped}
\nonumber \rho_A = \exp\left(-\sum_{j=0,1}^\infty \epsilon_j \hat n_j\right), \;\;\; \epsilon_j = 2 j \ln(J_{zz}+\sqrt{J_{zz}^2-1}),\\
\end{eqnarray}
where $\hat n_j$ is the fermion number operator, and the sum either starts from $j=0$ or from $j=1$. These two choices correspond to the cases with or without spontaneous symmetry breaking~\cite{Alba2017}, i.e., for a doubly degenerate ground state the sum starts with $j=0$ (and all levels are 2-fold degenerate), whereas for the symmetry breaking phase (single ground state), the sum starts from $j=1$. In both cases, the system is evidently free as $\hat n_j$ are just free-fermion operators.

We now test the asymptotically emergent free fermion behaviour dictated by Eq.~(\ref{eq:gapped}) by numerically calculating $D_{\cal F}$ for finite size systems. As explained in Sec.~\ref{sec:ext}, we use DMRG with open boundary conditions to obtain the entanglement spectrum of the ground state. Unless specified otherwise, we use bond dimension between 400-800 in order to converge the results.  As explained in Ref.~\onlinecite{LaflorencieBoundary}, the entanglement spectrum is different depending on whether the size of the subsystem is even or odd; in the rest of the paper, we focus on the cases where $L=4k$, i.e., the subsystem size, typically taken to be half of the total system size, contains an even number of sites, for which entropy is larger~\cite{LaflorencieBoundary}. 
\begin{figure}[htb]
    \centering
    \includegraphics[width=\linewidth]{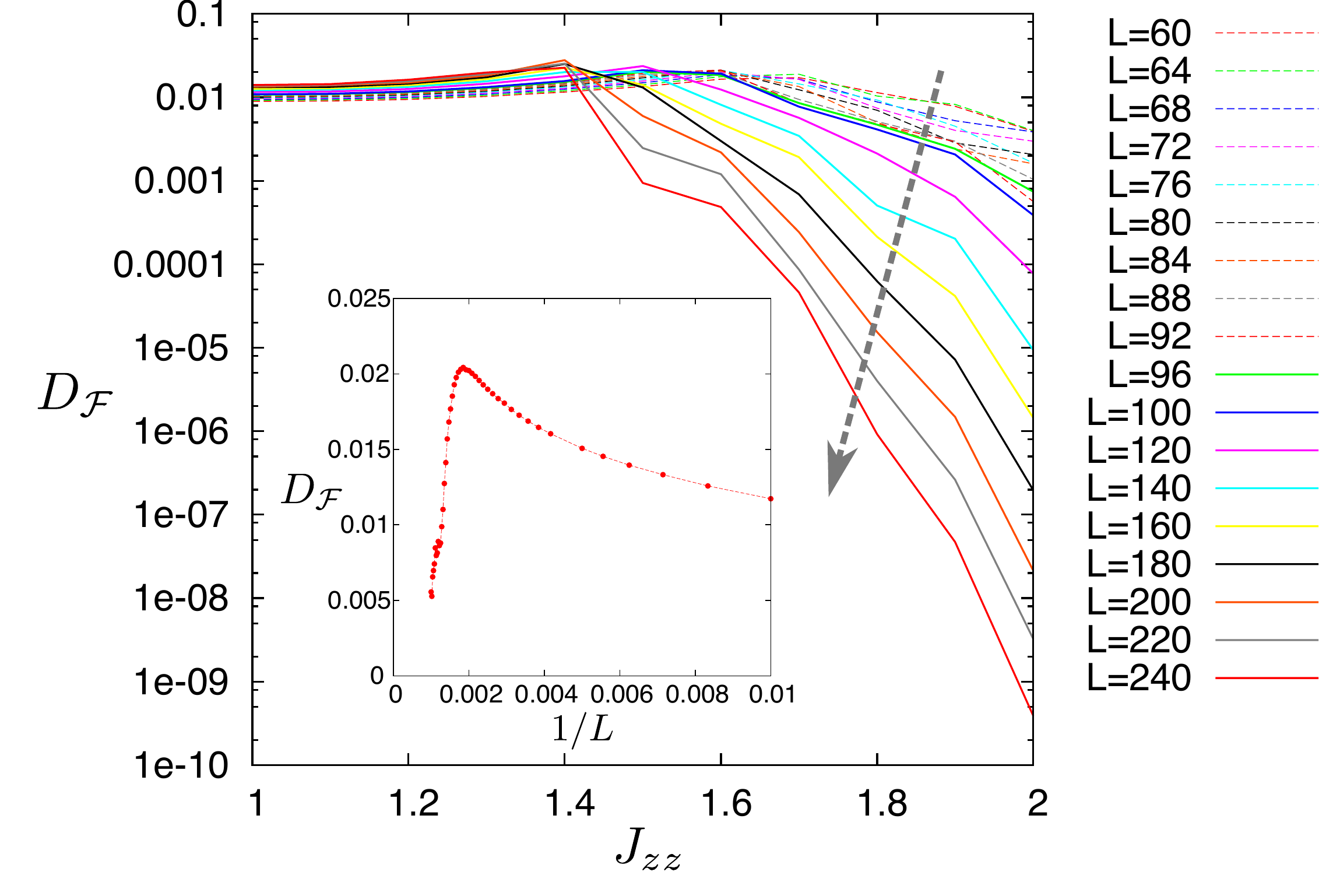}
    \caption{The interaction distance, $D_{\cal F} $, in the antiferromagnetic gapped phase of XXZ model with  $J_{zz}>1$, $J_{zz}'=0$, for various system sizes $L= 4k$. The arrow denotes the decreasing trend of $D_{\cal F} $ as the system size is increased. Inset shows the finite-size scaling of $D_{\cal F} $ for the fixed value $J_{zz}=1.2$ close to the transition. }
    \label{fig:gappedJzz}
\end{figure}

The interaction distance is shown in Fig.~\ref{fig:gappedJzz} for a range of $J_{zz}$ values in the gapped phase and different total system sizes $L$. From this figure we see that $D_{\cal F} $ appears to remain constant where $J_{zz}$ is close to 1, but then starts to exponentially decrease beyond some critical $J_{zz}^c$. The value of this $J_{zz}^c$ drifts to the left as the system size is increased, which suggests that in the thermodynamic limit, $D_{\cal F} $ will be zero for any $J_{zz}>1$. However, our results also illustrate that one may need to go to very large system sizes in order to start to see the free behaviour expected from results of Refs.~\onlinecite{PeschelIntegrable,Alba2017}. We see below that this behaviour is due to the  Berezinskii-Kosterlitz-Thouless (BKT) nature of the phase transition at $J_{zz}=1$~\cite{CazalillaRMP}. 

In order to test our previous interpretation of finite size scaling, in the inset of Fig.~\ref{fig:gappedJzz} we pick  $J_{zz}=1.2$, which belongs to the regime where $D_{\cal F} \approx const$ appears to hold, and then explicitly perform scaling with respect to system size.  We see that in small enough systems, $D_{\cal F} $ typically grows with system size -- this is the non-universal regime where the system is not big enough to accommodate the dressed DoF due to their large size $\ell$. For system sizes greater than some critical value $L_c$, which is itself a function of $J_{zz}$, $D_{\cal F} $ has opposite trend -- it decays with system size towards the zero value in an exponential fashion, as we expect from Refs.~\cite{PeschelIntegrable, Alba2017}. Thus, we confirm that the integrable XXZ model for $J_{zz}>1$ can be asymptotically described by free fermion in agreement with Eq.~(\ref{eq:gapped}). As an example, for $J_{zz}=5$ and for $L=400$ the lowest single particle energies $\epsilon_j$ are in agreement with the ones given by~(\ref{eq:gapped}) within an error of $10^{-6}$. Hence, the optimal free model obtained from the interaction distance faithfully identifies the result in Eq.~(\ref{eq:gapped}) obtained by integrable methods.

\subsection{Gapless Luttinger phase}

\begin{figure}[htb]
    \centering
    \includegraphics[width=\linewidth]{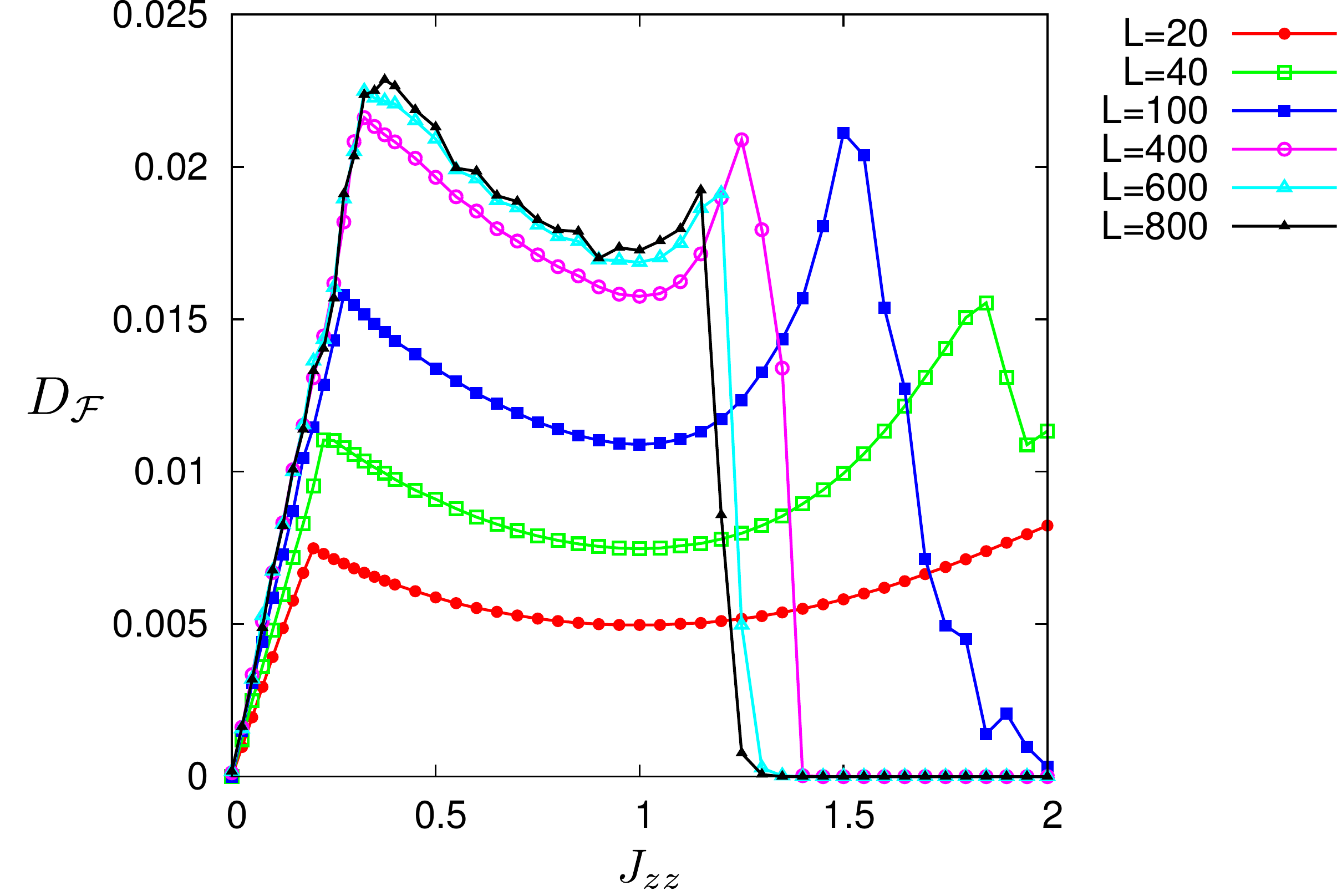}
    \caption{The interaction distance, $D_{\cal F} $, as a function of $J_{zz}$ spanning both gapped and gapless regimes. In the gapless regime for small $J_{zz}$, $D_{\cal F} $ exhibits a robust linear growth, $D_{\cal F} \propto J_{zz}$. Data is obtained by DMRG for a sequence of system sizes $L$ indicated in the legend.}
    \label{fig:edJzz}
\end{figure}

Next we move on to the gapless Luttinger liquid phase, which is realised in the XXZ model with $|J_{zz}|\leq 1$. For this coupling regime the XXZ model can also be solved via bosonisation~\cite{gogolin2004bosonization} that maps the low-energy behaviour of the system to that of a system of free bosons. Nevertheless, we want to continue our investigation in terms of the interaction distance and find the behaviour of $D_{\cal F}$ across the phase transition to the gapless regime of the Luttinger liquid. Similar to above, we first scan the behaviour of $D_{\cal F} $ as a function of $J_{zz}$ coupling across the range  $0\le J_{zz}\le2$ shown in Fig.~\ref{fig:edJzz}. First, we notice that in the gapped phase $J_{zz}>1$, the results clearly show the drift of $J_{zz}^c$ towards $J_{zz}=1$, as discussed previously in Fig.~\ref{fig:gappedJzz}. On the other hand, for $J_{zz} < 1$ we are in the Luttinger liquid phase. Intriguingly, we see that for $J_{zz} < 0.4$, $D_{\cal F} $ exhibits a robust, seemingly linear, growth with $J_{zz}$. Furthermore, the slope of the linear growth shows a weak dependence on system size $L$. More precisely, the slope depends on the size of the \emph{subsystem} $L_A$, which is fixed at $L_A=L/2$ in Fig.~\ref{fig:edJzz}.

A simple heuristic argument can explain the growth of $D_{\cal F} $. Since $D_{\cal F} $ is predominantly determined by the largest eigenvalues of $\rho_A$, or equivalently the lowest entanglement energies, it is important to know the low-lying structure of the entanglement spectrum in the Luttinger liquid phase. A general theorem by Bisognano and Wichmann~\cite{BisognanoWichmann1, BisognanoWichmann2}, applicable to systems described by relativistic quantum field theory, establishes a direct correspondence between the eigenvalues of $\rho_A$ and the energy eigenvalues of the Hamiltonian restricted to the subsystem~\cite{SwingleSenthill, GoldsteinSela, DalmonteBisognano}. Thus, for $|J_{zz}|<1$, the entanglement energies are given by the actual energies of a Luttinger liquid Hamiltonian for open boundary conditions. The Hamiltonian of a Luttinger liquid with open boundary condition is given, e.g., in Eq.(129) of Ref.~\onlinecite{CazalillaRMP} 
\begin{eqnarray}\label{eq:llham}
H = \sum_{q>0} \hbar v \; q \; a_q^\dagger a_q + \frac{\hbar \pi v}{2LK} (\hat N - N)^2,
\end{eqnarray}
where $v$ is the velocity, $K$ is the Luttinger parameter, $L$ denotes the system size, and $\hat N$ is the total number operator. For XXZ model, Bethe ansatz gives explicit expressions for $v$ and $K$~\cite{CazalillaRMP}:
\begin{eqnarray}\label{eq:Kv}
v &=& \frac{\pi v_{F}}{2} \frac{\sqrt{1-J_{zz}^2}}{\arccos J_{zz}},\\
K &=& \frac{1}{2-\frac{2}{\pi}\arccos J_{zz}},\nonumber
\end{eqnarray}
where $v_{F} = 1$ for $J=1$. Generally, $K$ and $v$ can be treated as phenomenological parameters. Since the Luttinger liquid Hamiltonian has a U(1) symmetry, the spectrum splits into different number sectors with a parabolic envelope given by the second term in the Hamiltonian. In addition, for fixed $N$, the first term in the Hamiltonian gives the spectrum of a chiral boson, with the tower of states whose degeneracies are equal to the number of partitions of an integer, i.e., the degeneracies are 1, 1, 2, 3, 5, 7, etc.

\begin{figure}[tb]
    \centering
    \includegraphics[width=\linewidth]{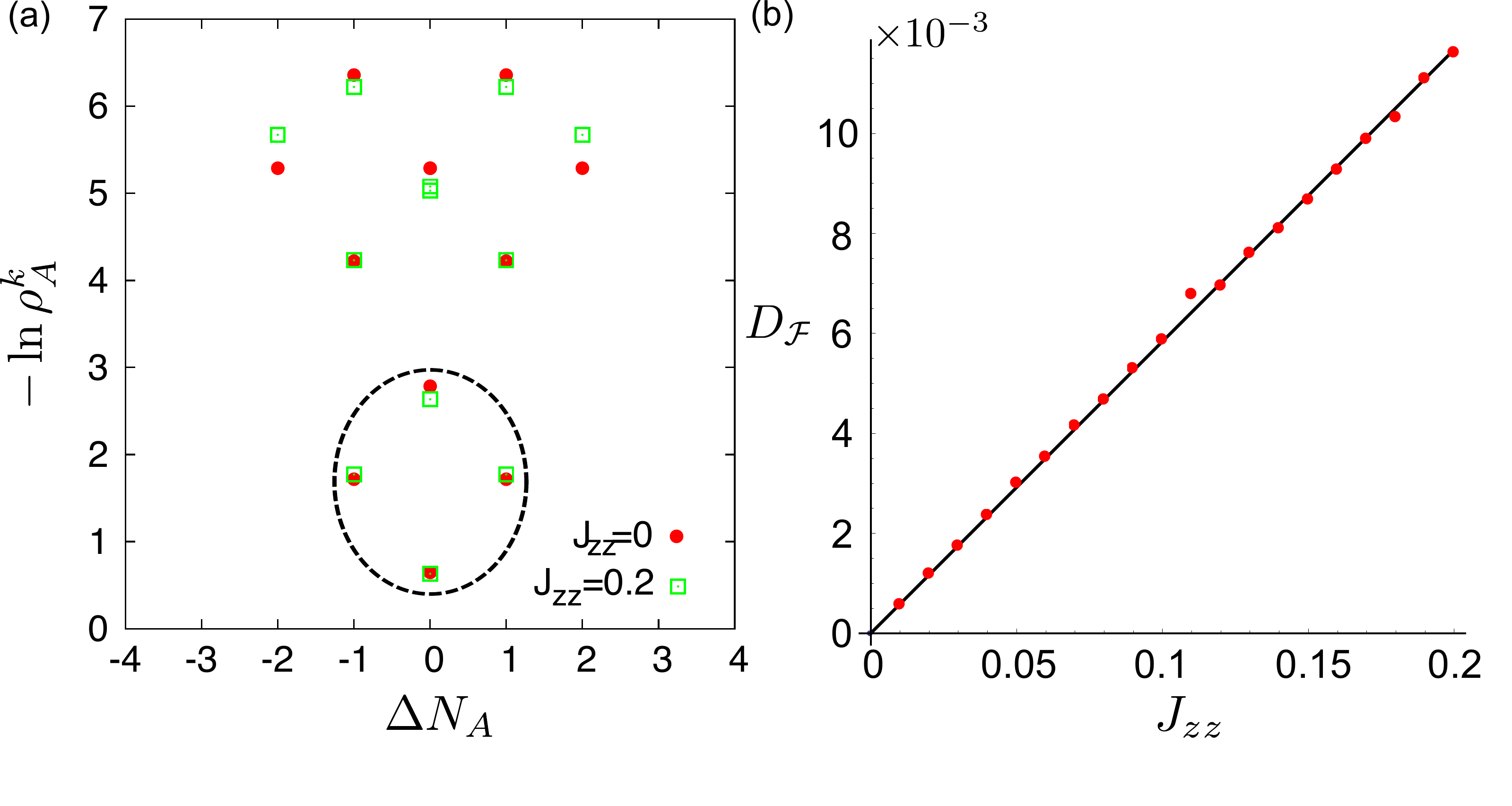}
    \caption{ (a) Entanglement spectrum of the ground state of XXZ model with $J_{zz}=0$ (XY model) and $J_{zz}=0.2$. Data is for system size $L=24$ obtained by exact diagonalisation. Entanglement energies, $-\ln\rho_k$, are plotted as a function of the number of particles in $A$ subsystem. The four lowest levels (indicated by the dashed circle) are responsible for the increase in $D_{\cal F} $ with $J_{zz}$. Comparing the case $J_{zz}=0$ with that of $J_{zz}=0.2$, we see that the symmetry of the four levels around the middle point [c.f. Eq.~(\ref{eqn:freespec})] gets destroyed in the presence of interactions, as the top most level slightly comes down, while the two degenerate ones move upwards. (b) The linear growth of $D_{\cal F} $ as a function of $J_{zz}$ obtained from DMRG is compared against the analytic ansatz in Eq.~(\ref{eq:dfapprox}), with $\ell_0\approx 4.6$ and system size $L=100$. 
    }
    \label{fig:es}
\end{figure}

The universality of the Luttinger liquid Hamiltonian implies that the general structure of its energy levels directly translates into the same structure of the \emph{entanglement energies} for the subsystem's reduced density matrix. The analysis of the entanglement spectrum from this point of view was performed in detail in Ref.~\onlinecite{LauchliCFT}, and we reproduce an example in Fig.~\ref{fig:es}(a) for a small XXZ periodic chain of $L=24$ spins with $J_{zz}=0$ and $J_{zz}=0.2$. The entanglement spectrum in Fig.~\ref{fig:es}(a) is plotted as a function of $\Delta N_A$, the relative number of particles in the subsystem $A$ compared to $N_A=N/2$. The spectrum splits into conformal towers corresponding to different particle numbers, $\Delta N_A = 0, \pm 1, \pm 2, \ldots$. The behaviour of $D_{\cal F} $ as a function of $J_{zz}$ can be explained by considering the lowest four entanglement energies, which have been indicated by a dashed circle in Fig.~\ref{fig:es}(a). From the form of the Hamiltonian in Eq.~(\ref{eq:llham}) (assuming Dirichlet boundary conditions), we see that these entanglement energies are given by
\begin{eqnarray}\label{eq:enten}
E_0 &=& e_0, \nonumber\\
E_1 &=& E_2 = e_0 \left( 1 + \frac{1}{2K}\right), \\
E_3 &=& 2 e_0.\nonumber
\end{eqnarray}
with $e_0 = \frac{\pi \hbar v}{L}$, i.e., two of them are in the sector with half the number of particles in $A$ ($\Delta N_A=0$), and two remaining energies are in number sectors that differ by $\pm 1$.

From Eq.~(\ref{eq:enten}), it is clear why $D_{\cal F}$ increases: when $K=1$ (XY model), the four levels are symmetric around the mid point  $(E_\mathrm{max}+E_\mathrm{min})/2$, thus describable by the free fermion modes as in Eq.~(\ref{eqn:freespec}) and $D_{\cal F} =0$; in all other cases, we obtain a set of four entanglement energies that are not symmetric around the middle point and thus cannot be described by~(\ref{eqn:freespec}). Hence, $D_{\cal F} $ is not zero and can be precisely quantified. For small $J_{zz}$ we can assume the corresponding free model is just the XY model, for which the (unnormalised) reduced density matrix eigenvalues are $\sigma_0 = 1$, $\sigma_1= \sigma_2= e^{-\beta_\mathrm{ent}/2}$, $\sigma_3= e^{-\beta_\mathrm{ent}}$. 
The interaction distance is then given by 
\begin{eqnarray}\label{eq:df2}
D_{\cal F}  = \frac{1}{2} \sum_{k=0}^3 |\frac{1}{Z} e^{-\beta_\mathrm{ent}E_k} - \frac{1}{Z_\sigma}\sigma_k|,
\end{eqnarray}
which can be evaluated using Eqs.~(\ref{eq:enten}). Note that the (inverse)  entanglement temperature $\beta_\mathrm{ent}$ should be set (for an open system) according to~\cite{LaflorencieRachel}
\begin{eqnarray}\label{eq:enttemp}
\beta_\mathrm{ent} = \frac{2\pi L_A}{v \ln(L_A/\ell_0)},
\end{eqnarray}
where $\ell_0$ is a lattice regularisation. The latter can be found, e.g., from the bipartite fluctuation of magnetisation~\cite{LaflorencieRachel}. Note that the final result for $D_{\cal F} $ involves products of the form $\beta_\mathrm{ent} E_k$. Hence, it continues to carry a weak subsystem-size dependence via the factor $\ln(L_A/\ell_0)$ from the definition of entanglement temperature. 

Using the expressions in Eqs.~(\ref{eq:df2}), (\ref{eq:enttemp}) and (\ref{eq:Kv}), and expanding to first order in $J_{zz}$, we obtain 
\begin{eqnarray}\label{eq:dfapprox}
D_{\cal F}  = J_{zz} \frac{3\pi\cosh(\pi^2/\ln \frac{L_A}{\ell_0})}{8 \ln\frac{L_A}{\ell_0} \cosh^4\left( \pi^2/(2\ln \frac{L_A}{\ell_0})\right)}   + \mathcal{O}(J_{zz}^2).\;\;\;\;\;\;
\end{eqnarray} 
Using e.g. $\ell_0 \approx 4.6$, this formula gives a good agreement against the $D_{\cal F} $ growth calculated in DMRG for a system of $L=100$ sites, as shown in Fig.~\ref{fig:es}(b). Hence, it is possible to faithfully determine the behaviour of $D_{\cal F}$ from a small number of lowest eigenvalues even if the system is in the gapless regime. We note that, unlike the gapped phase which was studied in Refs.~\onlinecite{PeschelIntegrable,Alba2017}, we are unaware of analytical results for the entanglement spectrum in the Luttinger liquid phase in the limit of an infinite system.

Finally, it is interesting to analyse what happens when our simple toy model in Eq.~(\ref{eq:enten}) is pushed beyond its validity when system size $L$ becomes very large. In this case, the structure of the entanglement spectrum will change, with an increasing number of entanglement levels becoming degenerate with each other. This is because of weak logarithmic dependence of entanglement temperature on the subsystem size, as noted above. Thus, based on our simple toy model, we may expect that there is a crossover between the linear increase of $D_{\cal F} $ to a decay in much larger system sizes. This investigation, however, is beyond the scope of present work.

\subsection{Probing the structure of dressed DoF}

From the previous analysis we have seen that the XXZ model has $D_{\cal F} \to 0$ as $L\to \infty$, for $J_{zz}>1$. This reflects the fact that the dressed DoF of the model tend to behave like free fermions even for finite but large system size, extending the infinite size result of Eq.~(\ref{eq:gapped}). Note that the internal structure of the dressed DoF is generated from the interactions due to a non-trivial rotation $Uc_jU^\dagger$. Nevertheless, for large enough system sizes and partitions $L_A$ the fermionic dressed DoF fit well within the region of the partition $L_A$ and its complement, as shown in Fig.~\ref{fig:qp}(a). Then the interaction distance measures the free particle correlations giving $D_{\cal F} \approx 0$, as shown in Fig.~\ref{fig:gappedJzz}. When $L_A$ becomes small compared to the size, $\ell$, of the dressed DoF, then the lowest part of the resulting entanglement spectrum will be influenced by the structure of the dressed DoF, as shown in Fig.~\ref{fig:qp}(b). The part of the entanglement spectrum that corresponds to the internal structure of the quasiparticles does not necessarily correspond to free correlations as the structure of $U$ comes from the presence of interactions. This is precisely the information about the dressed DoF that can be captured by $D_{\cal F}$.
\begin{figure}[htb]
    \centering
    \includegraphics[width=.95\linewidth]{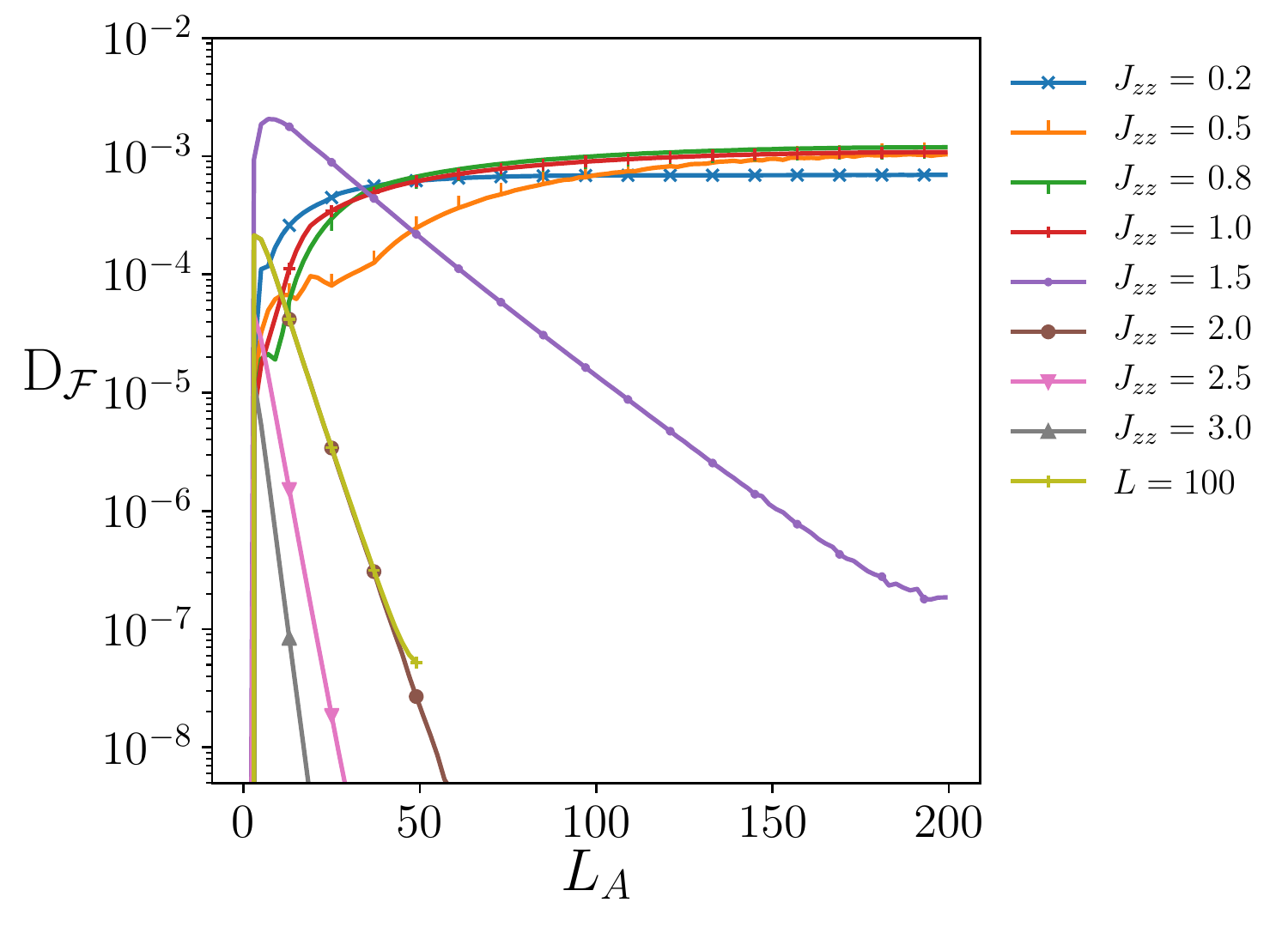}
    \caption{The interaction distance, $D_{\cal F} $, plotted on a logarithmic scale as a function of subsystem size $L_A$, on the integrable line $J_{zz}'=0$ for various $J_{zz}$. Total size of the chain is fixed at $L=400$ sites, with an additional data point at $L=100$ $J_{zz}=2$. We see that in the entire Luttinger liquid phase ($J_{zz}\leq 1$), there is no decay of $D_{\cal F} $ with subsystem size $L_A$. On the other hand, in the gapped antiferromagnetic phase, $D_{\cal F} $ decays exponentially with $L_A$, with an exponent that depends on $J_{zz}$. This decay reflects the exponential localisation of fermionic dressed DoF, that does not change with system size. 
    }
    \label{fig:dfcorrlen}
\end{figure}

\begin{figure}[htb]
	\centering
	\includegraphics[width=\linewidth]{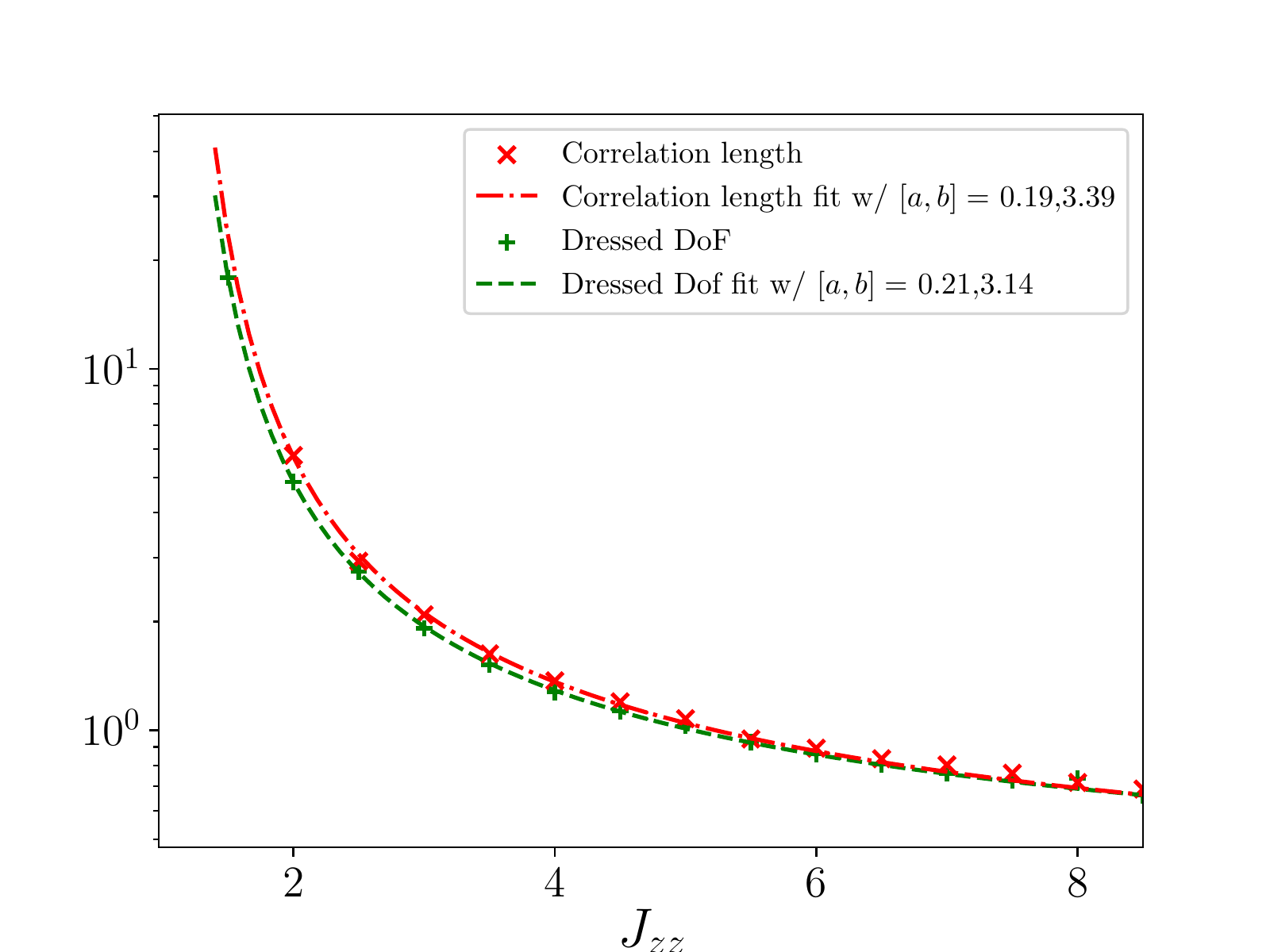}
	\caption{Numerically determined size $\ell$ of the dressed DoF and correlation length $\xi$ as a function of $J_{zz}$. Both data points are fit according to Eq.~(\ref{eq:bkt}), with the critical value $J_{zz}^c=1$. Both $\ell$ and the $\xi$ diverge with roughly the same functional form and $\ell\sim\xi$.
	}
	\label{fig:quasisize}
    \end{figure}

We can quantitatively extract the size of the dressed DoF of the XXZ model by examining the dependence of $D_{\cal F} $ on the subsystem size, $L_A$, for various choices $J_{zz}$ in the gapped and gapless regions. In this analysis we keep the total size of the system fixed at $L=400$. This size is large enough to be in the proper scaling regime giving $D_{\cal F} \to 0$ for many choices of $J_{zz} \gtrsim 1.5$. In Fig.~\ref{fig:dfcorrlen}, we vary the location of the partition for fixed system size. We see that there is an exponential increase in $D_{\cal F} $ when the subsystem size is reduced down to $L_A\sim 20$, then it goes identically to zero $D_{\cal F} =0$ for $L_A=1$, as calculated analytically~\cite{Turner}. This increase can be explained in terms of Fig.~\ref{fig:qp} that depicts localised dressed DoF. Assuming that the profile of the dressed DoF is exponential, we expect the scaling behaviour of $D_{\cal F} $ to be given by
\begin{eqnarray}
D_{\cal F}  \propto \exp(-L_A/\ell).
\label{eqn:size}
\end{eqnarray}
From this relation we can extract the size $\ell$ of the dressed DoF as a function of $J_{zz}$, as shown in Fig.~\ref{fig:quasisize}. We compare its divergence with the divergence of the correlation length $\xi$  as the system approaches the critical point at $J^c_{zz}=1$.  The correlation length is extracted from the spin-spin correlations 
\begin{eqnarray}
G(r)=\langle S^z_jS^z_{j+r}\rangle-\langle S^z_j\rangle\langle S^z_{j+r}\rangle,
\end{eqnarray}
using an exponential fit with a polynomial prefactor~\cite{Dugave_2015}
\begin{eqnarray}\label{eq:oz}
 G(r)\sim \frac{1}{r^2}e^{-r/\xi}.
\end{eqnarray}
It was recently pointed out~\cite{Rams} that using a simple exponential instead of the correct Ornstein-Zernike ansatz as in Eq.~(\ref{eq:oz}) would result in large errors for the estimated $\xi$. From the fit to Eq.~(\ref{eq:oz}), we find divergence of the correlation length as $J_{zz}$ approaches the critical point, as shown in Fig.~\ref{fig:quasisize}. The divergence takes the form characteristic of the BKT transition~\cite{CazalillaRMP}
\begin{eqnarray}\label{eq:bkt}
f(J_{zz}) = a\,\mathrm{exp}\left( \frac{b}{\sqrt{|J_{zz}-J^c_{zz}|}}\right),
\end{eqnarray}
with constant $a$ and $b$. Surprisingly, as we observe in Fig.~\ref{fig:quasisize}, $\ell$ also diverges with $J_{zz}$ in a similar fashion to $\xi$. This signals that in the present case they both depend on the energy gap of the system according to
\begin{eqnarray}
\ell\sim\xi\sim E_{\mathrm{gap}}^{-1}
\end{eqnarray}
within the gapped phase. The fitting parameters are $[ a,b ]=[ 0.19, 3.39 ],[ 0.21, 3.14 ]$ for $\xi$ and $\ell$, respectively, when fit to the critical point at $J^c_{zz}=1$. The value of $b$ for $\xi$ and $\ell$ are found to be within $3\%$ and $10\%$ of the asymptotic Bethe ansatz result~\cite{SchmitteckertExt}.

The exponential profile in $D_{\cal F} $ shows that the effective DoF are dressed by interactions with an exponential tail. As $J_{zz}$ is tuned away from $J_{zz}^c$ in the gapped phase, the dressed DoF move towards a single particle picture, with $J_{zz}\to\infty$ trivially free. Indeed, at that point one would find the interaction distance identically zero $D_{\cal F} =0$ for all choices of cut as it is possible, via a Jordan-Wigner transformation, to map the interacting Hamiltonian to one that is quadratic in its fermionic operators. Thus, the dressed DoF size $\ell$ is also trivially zero at this point. Within the Luttinger Liquid phase we see in Fig.~\ref{fig:dfcorrlen} no exponential profile attached to $D_{\cal F} $ as the partition is changed. So it is not possible to extract a meaningful, finite size $\ell$ due to the critical nature of this phase. 

\section{Quantifying interactions in the non-integrable regime}

Our previous analysis on the XXZ model complements the rigorously established results in the literature, obtained either by Bethe ansatz or bosonisation techniques, and calibrates $D_{\cal F}$ on an integrable model. Now we investigate the effect of next-nearest neighbour interactions, $J_{zz}'\neq 0$, which break the integrability of the model. 

\subsection{Interaction distance of the extended XXZ model}

Previous numerical studies of the extended XXZ model using DMRG~\cite{Mishra} have mapped out its phase diagram as a function of $J_{zz}$ and $J_{zz}'$. It was established that the phase diagram consists of four phases: the Luttinger liquid (LL) phase, two types of charge-density wave (CDW1, CDW2) phases, and a bond-ordered (BO) phase. 

\begin{figure}[htb]
    \centering
    \includegraphics[width=\linewidth]{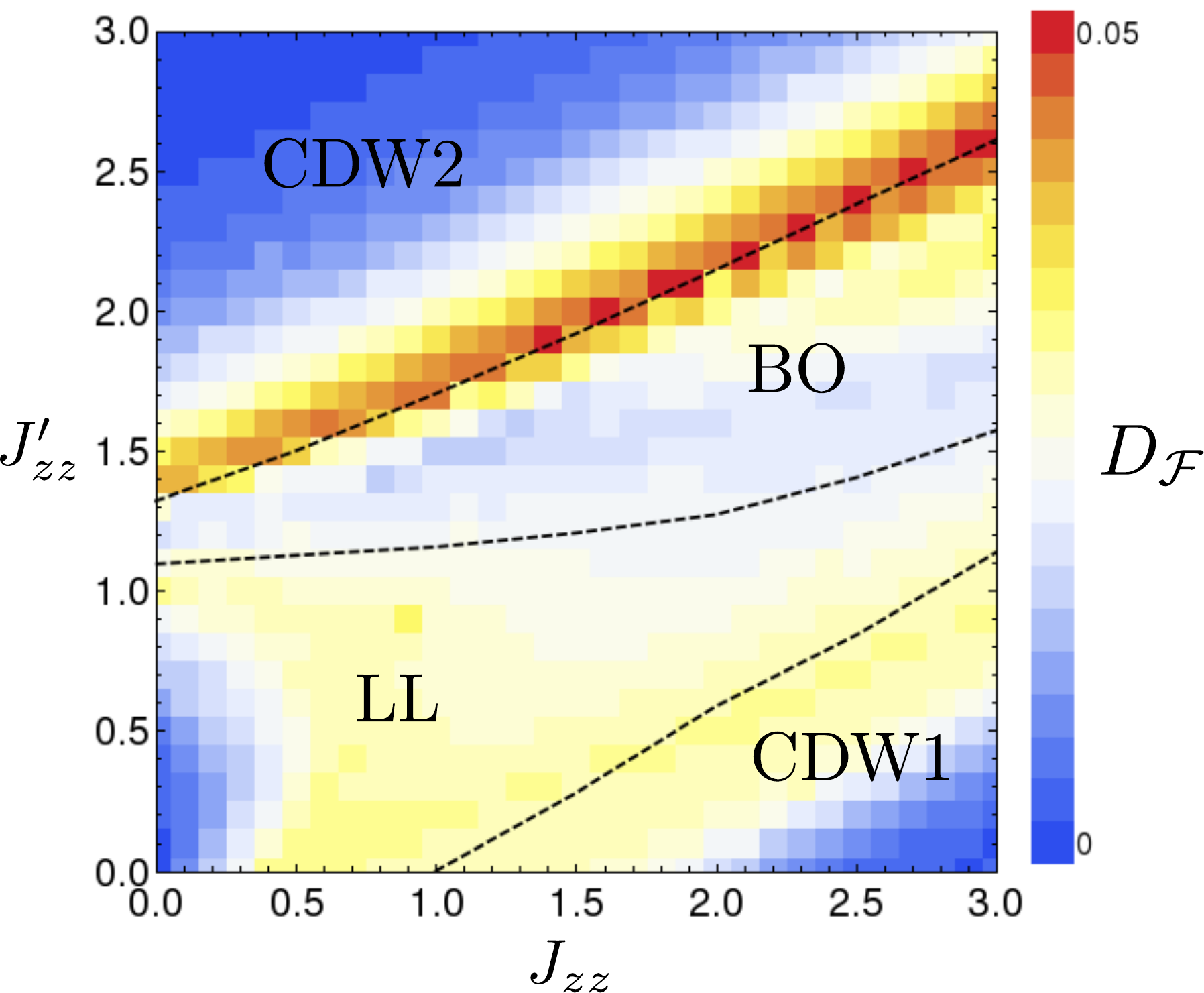}
    \caption{Interaction distance $D_{\cal F}$ (colour scale) across the 2D phase diagram $J_{zz}$-$J_{zz}'$ for system size $L=20$ with periodic boundary conditions, obtained by exact diagonalisation. Dashed lines are approximate phase boundaries reproduced from Ref.~\cite{Mishra}, which separate the following phases: gapless Luttinger liquid phase (LL), two types of charge density wave phases (CDW1 and CDW2), and a bond ordered phase (BO).}
    \label{fig:dfext}
\end{figure}
In Fig.~\ref{fig:dfext} we map out the phase diagram $J_{zz}$-$J_{zz}'$ based on the value of $D_{\cal F}$ in the ground state. Although the phase diagram in Fig.~\ref{fig:dfext} is obtained for a rather small system size ($L=20$, with periodic boundary conditions), its structure is broadly consistent with phase boundaries found in Ref.~\onlinecite{Mishra} indicated by dashed lines. In particular, the structure of $D_\mathcal{F}$ clearly reveals the presence of at least four different phases. The Luttinger liquid phase is dominated by the larger values of $D_{\cal F}$ as it corresponds to a gapless phase, compared to the gapped charge-density wave phases. Indeed, when the energy gap is small then the corresponding ground state is more susceptible to the presence of interactions and $D_\mathcal{F}$ is large. However, this is not true near the origin of the phase diagram where we see a semicircular lobe with small value of $D_\mathcal{F}$. On the integrable line $J_{zz}'=0$, this lobe corresponds to the regime of linear increase of $D_\mathcal{F}$ that we discussed in Fig.~\ref{fig:edJzz}, and we expect that a similar behaviour persists when a small amount of $J_{zz}'$ is added.

Beside the LL phase, there are also three ordered phases in the phase diagram in Fig.~\ref{fig:dfext}. The crystalline phases CDW1, CDW2 have a simple interpretation in the classical (``atomic") limit when the XY term in the Hamiltonian is completely switched off. In that limit, CDW1 is adiabatically connected to the degenerate N\'eel product states, $101010\ldots$ and $010101\ldots$, while CDW2 has a doubled unit cell, $110011001100\ldots$ (and translated copies). Finally, the BO phase~\cite{Hallberg} is defined by the finite value of the order parameter, $\langle \frac{1}{L}\sum_i (-1)^i (c_i^\dagger c_{i+1}+h.c.)\rangle$. All of these phases, being weakly correlated, are expected to have relatively low values of $D_\mathcal{F}$, as indeed confirmed by Fig.~\ref{fig:dfext}.

\begin{figure}[tb]
    \centering
    \includegraphics[width=\linewidth]{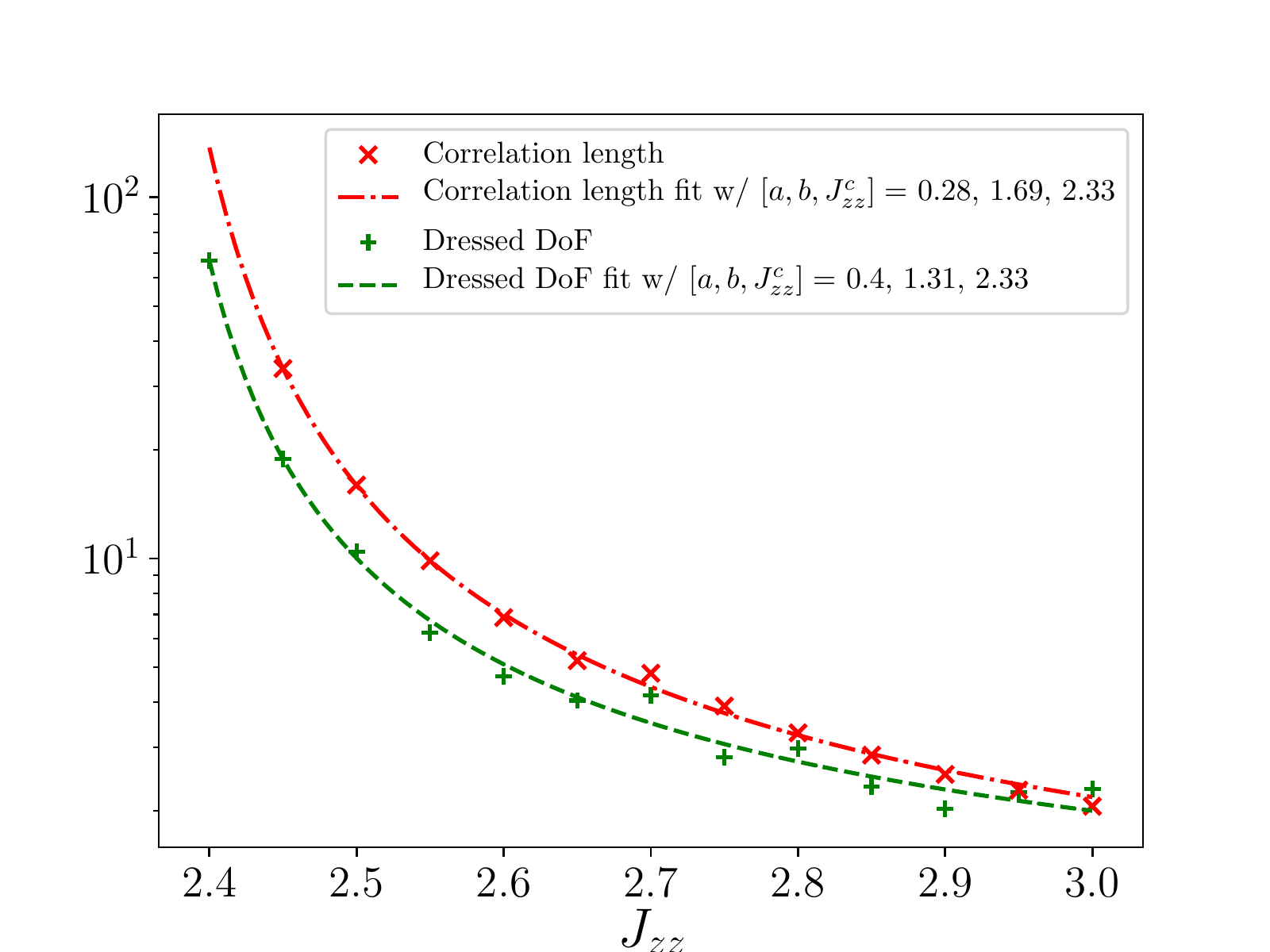}
    \caption{Numerically determined size $\ell$ of the dressed DoF and correlation length $\xi$ as a function of $J_{zz}$ for the line $J_{zz}' = 3-J_{zz}$, i.e., approaching the LL-CDW1 phase transition from within the gapped phase in Fig.~\ref{fig:dfext}. Data points are fit using the ansatz in Eq.~(\ref{eq:bkt}), with the critical value extracted as $J_{zz}^c=2.33$. Both $\ell$ and the $\xi$ diverge with the same functional form signalling that the phase transition LL-CDW1 in the extended model is also of BKT type.}
    \label{fig:extendedsize}
\end{figure}

We now investigate the scaling behaviour of $D_{\cal F}$ as we change the partition size $L_A$. From Eq.~(\ref{eqn:size}) we can extract the size $\ell$ of the dressed DoF as we did for the integrable XXZ model. Fig.~\ref{fig:extendedsize} shows that the size $\ell$ behaves similarly to the correlation length $\xi$ as it approaches the phase transition from the charge-density wave phase to the Luttinger liquid phase. Both $\ell$ and $\xi$ have a scaling behaviour consistent with BKT phase transition~\cite{SchmitteckertExt}, however there is a small deviation between their values in comparison to the integrable XXZ case.

\subsection{Integrability distance}

Finally, we are interested in the gapless Luttinger liquid phase of the extended XXZ model. As we have seen in Fig.~\ref{fig:dfext}, in small system sizes $D_\mathcal{F}$ has a non-monotonic behaviour in this phase, which motivates us to search for a more robust diagnostic.  The form of the extended XXZ Hamiltonian and our analysis so far suggest to introduce the following ``distance'', instead of $D_{\cal F}$, to quantitatively investigate the LL phase,
\begin{eqnarray}
D_{\mathrm{XXZ}} (\rho) = \min_{ -1\leq J_{zz} \leq 1} \frac{1}{2}\mathrm{tr}\sqrt{\Big( \rho- \sigma(J_{zz}) \Big)^2}\,.
\end{eqnarray}
Here $\rho$ is the reduced density matrix of the extended XXZ model and the minimisation is over $\sigma(J_{zz})$ which represents the reduced density matrix of the XXZ ground state at $-1\leq J_{zz} \leq 1$ ($J_{zz}'=0$). Unlike the definition of $D_{\cal F} $ in Eq.~(\ref{eqn:intdist}), note that $\sigma(J_{zz})$ is \emph{not} necessarily a free fermion density matrix, but that of the integrable XXZ model. Furthermore, we have used quotes in the name distance because $D_{\mathrm{XXZ}}$ characterises correlations in a single quantum state, rather than the spectral properties of the entire Hamiltonian (although the two should be linked in some way).
In general, a more appropriate quantity would be the \emph{integrability distance}, $D_{\cal I}$, defined as
\begin{eqnarray}\label{eq:di}
D_{\cal I} (\rho) = \min_{ \sigma\in {\cal I}} \frac{1}{2}\mathrm{tr}\sqrt{\left( \rho-\sigma \right)^2},
\end{eqnarray}
where the minimisation is performed over the set of all integrable models, ${\cal I}$. Several difficulties arise with this distance as explained below. For our immediate purposes, given the form of the extended XXZ Hamiltonian it seems natural to measure how far it is from the XXZ model in the Luttinger phase and hence to restrict our attention to $D_{\rm{XXZ}}$.

\begin{figure}[htb]
    \centering
    \includegraphics[width=\linewidth]{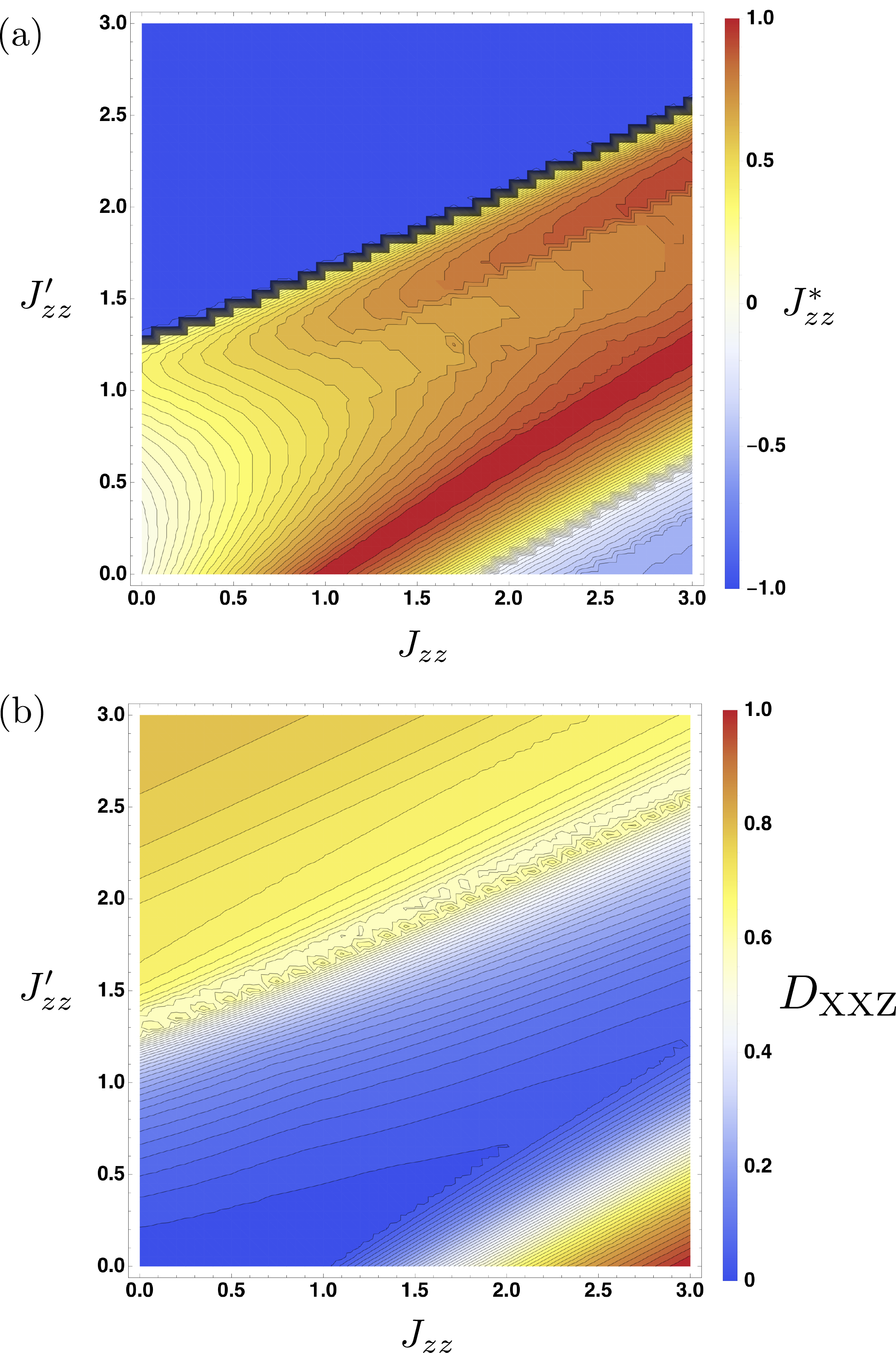}
    \caption{Integrability distance $D_{XXZ}$ across the two-dimensional phase diagram $J_{zz}$-$J_{zz}'$. Data is obtained by exact diagonalisation for system size $L=20$. Panel (a) shows the optimal integrable Luttinger coupling $J_{zz}^*$ for each point $(J_{zz},J_{zz}')$ in the phase diagram, with the corresponding minimal trace distance shown in panel (b). We see that the trace distance is small only in the diagonal strip of the phase diagram, which is consistent with the region identified as the Luttinger liquid phase in Ref.~\onlinecite{Mishra}.  
    }
    \label{fig:died}
\end{figure}

In practice, we evaluate $D_{\mathrm{XXZ}}$ numerically by precomputing $\sigma(J_{zz})$ for a dense set of values $J_{zz}$ distributed in the interval $\left[ -1, 1 \right]$. Then, for every value of the parameters $(J_{zz},J_{zz}')$, we obtain the ground state of the system, find its reduced density matrix $\rho$, and identify which of the precomputed $\sigma(J_{zz}^*)$ minimises the trace distance in Eq.~(\ref{eq:di}). This gives us two bits of information: (i) we obtain the optimal Luttinger liquid coupling $J_{zz}^*$ on the integrable line or, equivalently, the effective Luttinger parameter $K^*$ given by Eq.~(\ref{eq:Kv})~\cite{gogolin2004bosonization},  which best approximates the ground state at a \emph{non-integrable} point $(J_{zz},J_{zz}')$; (ii) we also obtain information about the quality of the approximation from the minimal achieved trace distance between $\rho$ and $\sigma(J_{zz}^*)$. If this minimal trace is not close to zero, the approximation is poor, and the description in terms of an integrable Luttinger liquid is not useful.

In Fig.~\ref{fig:died} we evaluate $D_{\mathrm{XXZ}}$ for the extended XXZ model by varying both interactions, $J_{zz}$ and $J_{zz}'$. Panel (a) shows the optimal integrable Luttinger coupling $J_{zz}^*$ for each point $(J_{zz},J_{zz}')$ in the phase diagram, with the corresponding minimal trace distance shown in panel (b). Indeed, from $J_{zz}^*$ the parameters $K$ and $v$ in (\ref{eq:Kv}) can be fully determined that give the Hamiltonian of the Luttinger liquid, Eq.~(\ref{eq:llham}), that best approximates the model at that point in the phase diagram. We see that the trace distance is small only in the diagonal strip of the phase diagram, which is consistent with the region identified as the Luttinger liquid phase in Ref.~\onlinecite{Mishra}.
Outside of this region, our optimal model is not accurate. Interestingly, along the phase boundary of the LL-CDW1 transtion, we find that the optimal model is $J_{zz}^*=1$. This is consistent with the literature~\cite{Mishra} which found that the Luttinger parameter $K$ along the entire phase boundary assumes the value $K^*=\frac{1}{2}$ (which translates to our $J_{zz}^*=1$).

\section{Conclusions and outlook}

In this article we qualitatively and quantitatively investigated the behaviour of the extended XXZ model focusing on the effect of interactions on different phases. This model has an integrable line that we probed with the interaction distance, $D_{\cal F}$. In the gapped regime we investigated the behaviour of its effective degrees of freedom, dressed by the interactions, that naturally emerge through the scaling analysis of the interaction distance. Hence, the interaction distance can efficiently describe the short and long distance behaviour of the model. Moreover, we provided analytical arguments about the behaviour of the interaction distance, $D_{\cal F}$, in the gapless Luttinger liquid regime of the integrable XXZ model. 

Outside the integrable line, a large part of the phase diagram of the extended XXZ model is expected to be described by the Luttinger liquid.   The investigation of this property motivates the introduction of the concept of integrability distance, $D_{\cal I}$ as in Eq.~(\ref{eq:di}). 
Similarly to the interaction distance, $\rho$ in Eq.~(\ref{eq:di}) can represent the Boltzmann-Gibbs density matrix of the system, or it can be the reduced density matrix of any eigenstate of the system, when it is bipartitioned. In this way $\rho$ can systematically probe all the relevant properties of integrable systems such as their energy spectrum and their quantum correlations.
Moreover, $D_{\cal I}$ could be directly expressed with respect to the eigenvalues of $\rho$ and $\sigma$. Nevertheless, two fundamental difficulties arise when working with Eq.~(\ref{eq:di}). One is that, unlike the free fermion density matrices, the general structure for $\sigma$ of integrable systems is not known. Another is that the set ${\cal I}$ of all possible integrable models one can envisage is not completely understood. In practice, one could simply list all the models that are known to be integrable up to now and run over this set. Due to the formidable complexity of varying over the whole space ${\cal I}$ of integrable models, we leave the investigation of the integrability distance, Eq.~(\ref{eq:di}), to future work. Identifying the general structure of $\sigma$ for (at least some) integrable models would be an important step. 
It would allow to evaluate $D_{\cal I}$ in full generality and thus help to quantitatively demonstrate how close non-integrable but physically relevant models are to mathematically idealised integrable examples. In view of these points, in our case, we restricted our attention to $D_{\mathrm{XXZ}}$ in order to quantitatively demonstrate that the extended XXZ model can be faithfully described by the Luttinger liquid with good accuracy. That being said, using $D_{\cal I}$ one could well imagine that at those points in the parameter space $J_{zz}$-$J_{zz}'$ where $D_{\mathrm{XXZ}}$ is not small,  the extended XXZ model is in fact ``closer to'' another integrable model.

\acknowledgements

This work was supported by the EPSRC grant EP/R020612/1. Statement of compliance with EPSRC policy framework on research data: This publication is theoretical work that does not require supporting research data.

 \bibliographystyle{apsrev4-1}
\bibliography{references}

\end{document}